\documentclass[12pt,reqno]{amsart}

\usepackage{amssymb,amsthm,amsmath,mathrsfs}
\usepackage{enumerate}
\usepackage{fullpage}
\usepackage{graphicx}
\usepackage{xcolor}

\begin{document}

\newtheorem{theorem}{Theorem}[section]
\newtheorem{lemma}[theorem]{Lemma}
\newtheorem{define}[theorem]{Definition}
\newtheorem{remark}[theorem]{Remark}
\newtheorem{corollary}[theorem]{Corollary}
\newtheorem{example}[theorem]{Example}
\newtheorem{assumption}[theorem]{Assumption}
\newtheorem{proposition}[theorem]{Proposition}
\newtheorem{conjecture}[theorem]{Conjecture}

\def\Ref#1{Ref.~\cite{#1}}
\def\Refs#1{Refs.~\cite{#1}}

\def\Rnum{{\mathbb R}}
\def\const{\text{const.}}
\def\smallbinom#1#2{{\textstyle \binom{#1}{#2}}}
\def\smallsum{\textstyle\sum}

\def\pr{{\rm pr}}
\def\X{\mathbf{X}}
\def\Y{\mathbf{Y}}
\def\id{\mathrm{id}}
\def\lieder#1{{\mathcal L}_{#1}}

\def\Jsp{\mathrm{J}^{(\infty)}}
\def\Esp{\mathcal{E}}

\def\Rop{{\mathcal R}}
\def\Dop{{\mathcal D}}
\def\Hop{{\mathcal H}}
\def\Jop{{\mathcal J}}
\def\p{{\mathrm p}}
\def\q{{\mathrm q}}
\def\a{\alpha}
\def\b{\beta}
\def\var{\text{var.}}
\def\dil{\text{dil.}}

\def\z{h}
\def\yone{A}
\def\ytwo{B}

\tolerance=50000
\allowdisplaybreaks[4]

\title{Conservation laws and exact solutions of\\a nonlinear acoustics equation\\by classical symmetry reduction}

\author{
Almudena P. M\'arquez${}^\dagger$,
Elena Recio,
Mar\'ia L. Gandarias
\\\\
\\
D\lowercase{\scshape{epartment}} \lowercase{\scshape{of}} M\lowercase{\scshape{athematics}}\\
U\lowercase{\scshape{niversity of}} C\lowercase{\scshape{adiz}}\\
11510 P\lowercase{\scshape{uerto}} R\lowercase{\scshape{eal}}, C\lowercase{\scshape{adiz}}, S\lowercase{\scshape{pain}}\\
}

\thanks{${}^\dagger$almudena.marquez@uca.es (A.P. Márquez)}

\begin{abstract}
Symmetries and conservation laws are studied for a generalized Westervelt equation 
which is a nonlinear partial differential equation
modelling the propagation of sound waves in a compressible medium.
This nonlinear wave equation is widely used in nonlinear acoustics
and it is especially important in biomedical applications such as
ultra-sound imaging in human tissue. 
Modern methods are applied to uncover point symmetries
and conservation laws that can lead to useful developments
concerning solutions and their properties.
A complete classification of point symmetries is shown
for the arbitrary function.
Local low-order conservation laws related to net mass of sound waves
are obtained by the multiplier method. 
Two potential systems are derived yielding 
potential symmetries and nonlocal conservation laws.
For the physical case interesting for this equation,
travelling wave solutions are studied leading to shock waves.
\end{abstract}

\maketitle

\section{Introduction}
Nonlinear and dissipative effects in the propagation of sound waves 
in a compressible medium \cite{HamBla-book}
is an interesting topic in the study of nonlinear wave equations.
Real-world applications are usually nonlinear problems.
For instance, relations between parameters like pressure,
density, temperature are nonlinear.
Additionally, it is important to consider dissipative effects
that widely appear in nonlinear acoustic waves.

An important biological/medical application is ultra-sound in human tissue \cite{MenCaiLiZHoNiuZhe,Sza,GueMarEglAliSer}.
Another examples of applications are
underwater imaging \cite{Lur},
parametric acoustic array in air \cite{GanYanKam},
musical acoustics of brass instruments \cite{Cam,MyePylGilCamChiLog},
chemistry and processing (sonochemistry)
including as major topics e.g.
chemical synthesis, environmental protection, electrochemistry \cite{MasLor},
and in physics,
quality control and material characterization \cite{MaeSev}.

A simple one-dimensional model is described 
by the Westervelt's equation \cite{Wes,Tar,Jor}
\begin{equation}\label{model}
(1 -2\a p)p_{tt} - \b p_{xxt} - 2\a p_t^2 = c^2 p_{xx}
\end{equation}
where $p(t,x)$ is the pressure fluctuation, 
$\b>0$ is the damping coefficient,
and $\a>0$ is the nonlinearity coefficient
arising from a quadratic equation of state \cite{EoS}
with the density $\rho(t,x)$ written in terms of the pressure,
\begin{equation}\label{eos}
\rho \approx p -\a p^2 -\b p_t.
\end{equation}

However,
it is also of interest to consider more general models 
including a generalized nonlinear term
and the damping term:
\begin{equation}\label{pde}
f(p)_{tt} - \b p_{xxt} = c^2 p_{xx}
\end{equation}
where the physical case will have $f(p)-p = h(p)$ being nonlinear ($h''(p)\neq0$),
following the structure of the original Westervelt's equation \eqref{model}.

For a dissipative version of the Westervelt's equation \eqref{model},
we previously studied 
symmetries, conservation laws, and its variational structure \cite{AncWes}. Moreover,
symmetries and conservation laws were recently found
for this dissipative Westervelt equation
in the case of spherical sound waves \cite{Anc2022a}
and also for a generalization \cite{Artur}.
These studies show the rich structure of this equation
and illustrate the numerous uses of modern symmetry analysis
for studying nonlinear wave equations.

Although the present work focuses on a one-dimensional equation,
it is of interest to consider as future work 
the two- or three-dimensional version.
Some works have already appeared in three spatial dimensions 
for the Westervelt's equation but from different points of view: 
\Refs{SolSheThi,ManSolShe} in numerical solutions,
\Ref{Kal} in the initial-value problem analysis,
and \Ref{Chi} in the existence and uniqueness of solutions.
The purpose of the present work is to study symmetries, conservation laws, and exact solutions in the general case \eqref{pde}, 
which has not been previously treated in the literature.

An important aspect of any wave equation is its symmetries and conservation laws.
Noether (1918) first derived conserved quantities from symmetries. 
Her theorem relates the fundamental relation
between the symmetries of a physical system
and the conservation laws,
and motivated physicists to study symmetries of physical systems.
This equation does not have a local Lagrangian formulation in terms of $p$,
thus Noether's theorem is inapplicable.
Instead,
its modern generalization using multipliers is applied to determine the local low-order conservation laws.

Beyond being basic structures,
they have several physical and mathematical applications:
group-invariant exact solutions; 
mapping of given solutions into new solutions;
linearizing transformations and integrability structures; 
balance equations for physical conserved quantities; 
conserved norms needed in analysis of solutions; 
accuracy checks of numerical solutions; 
and discretizations with inherent good properties.

Particularly,
this paper is devoted to obtaining the following results for the generalized Westervelt equation \eqref{pde}:
symmetries and low-order conservation laws; new conserved quantities; potential symmetries and nonlocal conservation laws arising from potential systems; shock wave solutions.

In Section \ref{sec:symms},
all point symmetries will be shown with their commutator structure and physical meaning.

In Section \ref{sec:conslaws},
all local conservation laws (of low-order) will be determined by using
the multiplier method which provides a modern form of the Noether's theorem for non-variational PDEs. 
These conservation laws turn out to depend on
an arbitrary function
and are shown to be related to mass conservation. 

In Section \ref{sec:potsys},
two potential systems will be derived 
and, for both, a first- and a second-layer potential are introduced. 
The point symmetries and low-order conservation laws are obtained for both layers of both potential systems.
These results include potential symmetries and nonlocal conservation laws.
This is a common way of finding symmetries and conservation laws
admitted by the equation that are not inherited from any
of the local point symmetries and local conservation laws 
of the governing equation.

In Section \ref{sec:tws},
travelling wave solutions of
the physical case of equation \eqref{pde}
with $f(p)-p=h(p)$ being nonlinear will be studied.
Specifically, interesting shock waves are obtained.

In Section \ref{sec:conclusions},
some conclusions are given.

Throughout, we work in the setting of jet space,
with tools from variational calculus.
See \Refs{Ovs-book,Olv-book,BCA-book,Anc-review} for the general theory of 
symmetries, conservation laws, and variational structures for PDEs.

\section{Point symmetries}
\label{sec:symms}

Symmetries are a fundamental structure of nonlinear PDEs. 
They yield transformation groups that map
the set of solutions of a given PDE into itself and thus
can be used to find group-invariant solutions. 
See \Refs{Olv-book,BCA-book} for a general discussion
of symmetries and their applications to PDEs.

A Lie point symmetry is a continuous transformation group
on the space of independent and dependent variables $(t,x,p)$
such that the equation is invariant. 
The infinitesimal form of the transformation group consists of
infinitesimal transformations on $(t,x,p)$ given by 
\begin{equation}\label{infinitesimal.point.transf}
t \to t +\epsilon \tau(t,x,p) + O(\epsilon^2), 
\quad
x \to x +\epsilon \xi(t,x,p) + O(\epsilon^2), 
\quad
p \to p +\epsilon \eta(t,x,p) + O(\epsilon^2)
\end{equation}
where $\epsilon\in\Rnum$ is the group parameter. 
The condition of invariance of the generalized Westervelt equation \eqref{pde} is given by 
\begin{equation}\label{inv.cond}
\big(\pr\X( f(p)_{tt} - \b p_{txx} - c^2 p_{xx} )\big)\big|_\Esp =0
\end{equation}
where 
\begin{equation}\label{X.generator}
\X=\tau(t,x,p)\partial_t + \xi(t,x,p)\partial_x + \eta(t,x,p)\partial_p
\end{equation}
is the operator generating an infinitesimal point transformation \eqref{infinitesimal.point.transf}; 
$\pr$ denotes prolongation of $\X$ to derivatives of $p$;
and $\Esp$ denotes the solution space of equation \eqref{pde}.

The invariance condition \eqref{inv.cond} is called the symmetry determining equation. This determining equation splits
with respect to derivatives of $p$, thereby yielding an overdetermined linear system of PDEs for the functions in the symmetry generator \eqref{X.generator}. 

A classification of the admitted point symmetries is given by solving the determining system for $\tau$, $\xi$, $\eta$, along with $f''(p)\neq 0$ and $\b\neq0$. 
This straightforwardly determines the solution and yields the following classification result.

\begin{theorem}
\label{thm:pointsymms}

The point symmetries of the generalized Westervelt equation \eqref{pde} with $f''(p)\neq 0$ and $\b\neq0$ are comprised by
a time-translation and a space-translation
\begin{equation}\label{transl}
\X_1 = \partial_t,
\quad
\X_2 = \partial_x .
\end{equation}
For 
$f(p)=k(p+p_0)^{1+q}$,
an additional
scaling in $x$ and $p$
\begin{equation}\label{scal1}
\X_3 = q x \partial_x - 2 (p+p_0) \partial_p.
\end{equation}
For
$f(p)=\tfrac{k}{(p+p_0)^3}$,
an additional
non-rigid scaling in $x$ and $p$
\begin{equation}\label{scal2}
\X_4 = x^2 \partial_x + x (p+p_0) \partial_p .
\end{equation}
\end{theorem}

The symmetry generators have the following commutator structure:
the two translations \eqref{transl} commute; 
the two scalings \eqref{scal1} and \eqref{scal2} commute with the time-translation;
the first scaling \eqref{scal1} acts as a multiple of the identity on the space-translation
and on the second scaling \eqref{scal2}.
The non-zero commutators in this structure are given by 
\begin{subequations}\label{symm.alg}
\begin{align}
& 
[\X_2,\X_3] = q \X_2 ,
\\
&
[\X_2,\X_4] = -\tfrac{1}{2} \X_3 |_{q=-4}, 
\\
&   
[\X_3,\X_4] = -4 \X_4 .
\end{align}  
\end{subequations}

There is an alternative way to formulate infinitesimal symmetries,
which will be useful for comparison in the following sections. 

A generator \eqref{X.generator} of an infinitesimal point transformation
on the variables $(t,x,p)$
has an equivalent evolutionary form in which $(t,x)$ are invariant,
which is called the characteristic form of $\X$ and is given by 
\begin{equation}\label{P.generator}
\hat\X=P\partial_p,
\quad
P = \eta(t,x,p) - \tau(t,x,p) p_t - \xi(t,x,p) p_x.
\end{equation}
Function $P$ is named the characteristic of the generator.
The equivalence is readily understood
from the action of the generator on functions $p$. 
Specifically, the action is given by the mapping
\begin{equation}\label{p.mapping}
p=f(t,x) \to p^{(\epsilon)} = f(t,x) + \epsilon P|_{p=f(t,x)} + O(\epsilon^2).
\end{equation}

For the four symmetries \eqref{transl}--\eqref{scal2}, 
\begin{equation}\label{P.symms}
\begin{aligned}
&
P_1 = -p_t ,
\quad
P_2 = -p_x,
\quad
P_3 = - 2 (p+p_0) - q  x p_x,
\quad
P_4 = x (p+p_0) - x^2 p_x 
\end{aligned}
\end{equation}
are their characteristic form. 
The symmetry determining equation \eqref{inv.cond} has an equivalent formulation
directly depending on function $P$
under the mapping \eqref{p.mapping}:
\begin{equation}\label{symm.deteqn}
\big(\pr\hat\X( f(p)_{tt} - \b p_{txx} - c^2 p_{xx} )\big)\big|_\Esp 
= \big( D_t^2 (f(P)) -\b D_t D_x^2 P  - c^2 D_x^2 P \big)\big|_\Esp =0 
\end{equation}
where $D_t$ and $D_x$ denote total deritatives which 
commute with $\pr\hat{\X}$. 
Unlike partial derivatives,
total derivatives approximate the function with
respect to all of its arguments, not only a single one.
This is the characteristic form of the symmetry determining equation.
It can be solved in the same way as the canonical determining equation
but it has the advantage that the prolongation of $\hat\X$ is not needed explicitly because of the property that it commutes with total derivatives. 

\section{Low-order conservation laws}
\label{sec:conslaws}

A conservation law of equation \eqref{pde} is a continuity equation
\begin{equation}\label{conslaw}
(D_t T + D_x \Phi)|_\Esp =0
\end{equation}
holding on the solution space $\Esp$ of equation \eqref{pde},
where $T$ is the conserved density and $\Phi$ is the spatial flux
and both are functions depending on $t$, $x$, $p$, and derivatives of $p$.

Integration of a continuity equation \eqref{conslaw} over the spatial domain
$\Omega\subseteq \Rnum$ leads to a conserved integral
\begin{equation}\label{cons.integral}
C=\int_{\Omega} T\,dx\big|_\Esp
\end{equation}
where
\begin{equation}
\frac{dC}{dt} = -\Phi|_{\partial\Omega}\big|_\Esp
\end{equation}
stating that the rate of change of the integral quantity \eqref{cons.integral}
is equal to the net spatial flux leaving the domain $\Omega$
over the domain boundary $\partial\Omega$. 
For solutions $p(t,x)$ of equation \eqref{pde} whose net flux vanishes, the integral $C$ will be conserved
(namely, time-independent). 

If $T=D_x\Theta$ and $\Phi=-D_t\Theta$ hold for all solutions,
where $\Theta$ is a function depending on $t$, $x$, $p$, and derivatives of $p$,
then the continuity equation contains no useful information about solutions $p(t,x)$.
In this situation, the conservation law is called trivial.
If two conservation laws differ by a trivial conservation law,
then they are considered to be locally equivalent since they contain the same information about solutions $p(t,x)$ on the domain $\Omega$.

For the purpose of computing conservation laws,
the simplest way is to use the multiplier method \cite{AncBlu2002b}
since there is a general connection between classifying all conserved quantities of a given form
and classifying all multipliers of an associated form.
For instance,
all conserved quantities of energy-momentum type correspond to
linear multipliers (first-order in derivatives of the dependent variable).

A multiplier for equation \eqref{pde} is a function depending on $t$, $x$, $p$, and derivatives of $p$,
such that it is non-singular on solutions of equation \eqref{pde} and its product with the equation is equal to a space-time total divergence 
\begin{equation}\label{multreqn}
(f(p)_{tt} - \b p_{txx} - c^2 p_{xx})Q= D_t T + D_x \Phi
\end{equation}
for some $T$ and $\Phi$.

There is an explicit relationship between conservation laws (up to local equivalence) and multipliers.           
To determine the multipliers, it is useful to apply the Euler operator
with respect to $p$ to the multiplier equation \eqref{multreqn}:
\begin{equation}
E_p( (f(p)_{tt} - \b p_{txx} - c^2 p_{xx})Q ) =0
\end{equation}
holding identically.
This condition can be split into an overdetermined linear system for $Q$, $f(p)$, $\b$,
which is similar to the system for point symmetries in characteristic form in terms of $P$.

In general,
conservation laws for fundamental physical quantities such as momentum and energy
arise from multipliers of lower order than the order of the main equation \cite{Anc-review}. 
Such low-order multipliers for equation \eqref{pde} are of the form
$Q(t,x,p,p_t,p_x,p_{tt},p_{tx},p_{xx})$. 
The overdetermined linear system can be directly solved determining the following multipliers.

\begin{proposition}
The low-order multipliers for the generalized Westervelt equation \eqref{pde} with $f''(p)\neq0$ and $\b\neq0$ are
\begin{equation}
Q_1 =1,
\quad
Q_2 =t ,
\quad
Q_3=x,
\quad
Q_4=tx .
\end{equation}
\end{proposition}

The conservation law can be obtained from a multiplier
by several methods explained in \Ref{Anc-review}.
Then, we get to the following result.

\begin{theorem}
The low-order conservation laws for the generalized Westervelt equation \eqref{pde} with $f''(p)\neq0$ and $\b\neq0$ are
\begin{align}
& 
T_1 = f(p)_t  
, 
&&
\Phi_1 = - c^2 p_x - \b p_{tx}
 ,
 \label{conslaw1}
\\
& 
T_2 =  - f(p) + t f(p)_t 
, 
&&
\Phi_2 =  - t (c^2 p_x + \b p_{tx} ) ,
 \label{conslaw2}
\\
& 
T_3 = x f(p)_t , 
&&
\Phi_3 =    c^2 ( p - x p_x ) +\b (p_t- x p_{tx})
,
 \label{conslaw3}
\\
& 
T_4 = -x( f(p) - t f(p)_t ) 
 , 
&&
\Phi_4 =  c^2 t (  p -  x p_x ) +\b t (p_t- x p_{tx}) .
 \label{conslaw4}
\end{align} 
\end{theorem}

The conserved quantities coming from these conservation laws 
defined on $\Omega = (-\infty,\infty)$ 
will now be discussed. 

\subsection{Conserved quantities}

The conserved integrals arising from the conservation laws \eqref{conslaw1} and \eqref{conslaw3} are
\begin{equation}\label{C1}
C_1 = \int_{-\infty}^{\infty} f(p)_t \, dx
\end{equation}
and
\begin{equation}\label{C3}
C_3 = \int_{-\infty}^{\infty} x f(p)_t \, dx.
\end{equation}
For their physical interpretation is important the relation 
between the pressure $p(t,x)$ and the density $\rho(t,x)$
via the equation of state \eqref{eos},
then we can write
$\rho = g(p)$
for some function $g$.
Thus,
the conserved integrals \eqref{C1} and \eqref{C3} can be written in terms of the density $\rho(t,x)$,
\begin{equation}\label{C1.dens}
C_1 = \int_{-\infty}^{\infty} f(g^{-1}(\rho))_t \, dx = \tfrac{d}{dt} f(m(t))
\end{equation}
and 
\begin{equation}\label{C3.dens}
C_3 = \int_{-\infty}^{\infty} x f(g^{-1}(\rho))_t \, dx = \tfrac{d}{dt} f(m^x(t)),
\end{equation}
where $m(t)=\int_{-\infty}^{\infty} f(g^{-1}(\rho)) \, dx$ and $m^x(t)=\int_{-\infty}^{\infty} x f(g^{-1}(\rho)) \, dx$
respectively correspond to a generalized net mass and a generalized $x$-weighted net mass.

Conservation laws \eqref{conslaw2} and \eqref{conslaw4} yield the conserved integrals
\begin{equation}\label{C2.dens}
C_2 = \int_{-\infty}^{\infty} (-f(g^{-1}(\rho)) + t f(g^{-1}(\rho))_t )\, dx = t \tfrac{d}{dt} f(m(t)) -  f(m(t))
\end{equation}
and
\begin{equation}\label{C4.dens}
C_4 = \int_{-\infty}^{\infty} -x (f(g^{-1}(\rho)) - t f(g^{-1}(\rho))_t )\, dx = t \tfrac{d}{dt} f(m^x(t)) -  f(m^x(t)) .
\end{equation}
When $t=0$,
$-C_2|_{t=0}=f(m(0))$ and $-C_4|_{t=0}=f(m^x(0))$,
which respectively correspond to the initial generalized net mass
and the initial generalized $x$-weighted net mass.

\section{Potential systems, symmetries and conservation laws}
\label{sec:potsys}

As is well known \cite{BCA-book,AncBlu2002b},
equations can possess nonlocal conservation laws yielding
conserved integrals apart from the ones that arise from local conservation laws.
A standard way of finding nonlocal conservation laws is through 
the introduction of a potential variable,
which typically is obtained by writing 
the conserved density and spatial flux in a local conservation law
in a trivial (nonlocal) form.

\subsection{First potential system}

Since the generalized Westervelt equation \eqref{pde}
has the form of a conservation law \eqref{conslaw}, 
a potential $u(t,x)$ can be introduced
yielding the first-layer potential system
\begin{subequations}
\begin{align}
f'(p) p_t - \mu p_{xx} & = u_x,
\label{1stlayer.ux}
\\
c^2 p_x + (\b - \mu) p_{tx} & = u_t .
\label{1stlayer.ut}
\end{align}
\end{subequations}
From equations \eqref{1stlayer.ux}--\eqref{1stlayer.ut},
two second-layer potentials $v(t,x)$, $w(t,x)$ can be introduced
yielding the second-layer potential system
\begin{subequations}
\begin{align}
f(p) =  v_x, 
\quad
&
\mu p_x + u = v_t ,
\label{2ndlayer.vx} 
\\
c^2 p + (\b - \mu - \sigma) p_t =  w_t , 
\quad
&
-\sigma p_x + u = w_x .
\label{2ndlayer.wx}
\end{align}
\end{subequations}
For these first- and second-layer potential systems
two free constants $\mu$ and $\sigma$ 
have been introduced, respectively,
to seek the most general result.

\subsubsection{First-layer potential point symmetries}

Potential symmetries in characteristic form 
\begin{equation}\label{pot1.1stlayer.X}
\hat\X = P^p\partial_p + P^u\partial_u
\end{equation}
arise from the equations
\begin{subequations}
\begin{align}
D_t (P^p f'(p))- \mu D_x^2 P^p - D_x P^u |_\Esp& = 0,
\label{pot1.1stlayer.det.symm.1}
\\
c^2 D_x P^p + (\b - \mu) D_t D_x P^p - D_t P^u|_\Esp& = 0 
\label{pot1.1stlayer.det.symm.2}
\end{align}
\end{subequations}
where $\Esp$ denotes the solution space of equations \eqref{1stlayer.ux}--\eqref{1stlayer.ut}.

The determining equations \eqref{pot1.1stlayer.det.symm.1}--\eqref{pot1.1stlayer.det.symm.2}
can be directly solved obtaining all point symmetries of the first-layer potential system \eqref{1stlayer.ux}--\eqref{1stlayer.ut}
with
\begin{subequations}
\begin{align}
P^p&=\eta^p(t,x,p,u)-\tau(t,x,p,u)p_t-\xi(t,x,p,u)p_x,
\\
P^u&=\eta^u(t,x,p,u)-\tau(t,x,p,u)u_t-\xi(t,x,p,u)u_x
\end{align}
\end{subequations}
as their characteristic form.

\begin{theorem}\label{thm:pot1.1stlayer.symms}
The point symmetries of the first-layer potential system \eqref{1stlayer.ux}--\eqref{1stlayer.ut} with $f''(p)\neq0$ and $\b\neq0$ are comprised by a time-translation, a space-translation, and a $u$-translation.
Their characteristic forms are
\begin{equation}
P_1^u = -u_t ,
\quad
P_2^u = -u_x,
\quad
P_3^u = 1.
\end{equation}
For
$f(p)=k(p+p_0)^{1+q}$,
an additional scaling in $x$, $p$, $u$
\begin{equation}
P_4^u = -(q+2)u - q x u_x .
\end{equation}
\end{theorem}

Application of \eqref{pot1.1stlayer.X} 
to equations \eqref{1stlayer.ux}--\eqref{1stlayer.ut} leads to
$f'(p)_tP^p + f'(p)D_tP^p - \mu D_x^2 P^p = D_xP^u$
and $c^2D_x P^p +(\b -\mu)D_tD_x P^p = D_t P^u$.
Solving the second equation for $P^p$ gives
$(\beta-\mu)D_t P^p = - c^2 P^p + D_x^{-1}D_t P^u$
and substituting into the first equation yields
$f'(p)_tP^p  - \frac{c^2}{\b -\mu}f'(p) P^p - \mu D_x^2 P^p = 
D_xP^u - \frac{1}{\b -\mu}f'(p) D_x^{-1}D_t P^u$.
Then, writing it in operator form acting on $P^p$ we have
$( f'(p)_t - \frac{c^2}{\b -\mu}f'(p) - \mu D_x^2 ) P^p = 
D_xP^u - \frac{1}{\b -\mu}f'(p) D_x^{-1}D_t P^u$
and inverting the operator,
\begin{equation}\label{1stlayer.symm.project}
P^p = ( f'(p)_t - \frac{c^2}{\b -\mu}f'(p) - \mu D_x^2 )^{-1} ( D_xP^u - \frac{1}{\b -\mu}f'(p) D_x^{-1}D_t P^u )
\end{equation}
yields the projection from the first-layer layer potential system \eqref{1stlayer.ux}--\eqref{1stlayer.ut}
to the PDE \eqref{pde}.


Substitution of the four symmetries from Theorem \ref{thm:pot1.1stlayer.symms}
into projection \eqref{1stlayer.symm.project} yields
\begin{equation}
\begin{aligned}
&
P_1^p = -p_t ,
\quad
P_2^p = -p_x,
\quad
P_3^p = 0,
\quad
P_4^p = -2(p+p_0) - q x p_x.
\end{aligned}
\end{equation}
The first two symmetries and the last symmetry are inherited from equation \eqref{pde} via the projection \eqref{1stlayer.symm.project}.
The other symmetry only exists for the first-layer potential system \eqref{1stlayer.ux}--\eqref{1stlayer.ut}. 

The correspondence between the inherited point symmetries 
of the potential system \eqref{1stlayer.ux}--\eqref{1stlayer.ut} and the point symmetries of equation \eqref{pde}
is
\begin{equation}
P^u_1 \leftrightarrow P_1;
\quad
P^u_2 \leftrightarrow P_2;
\quad
P^u_4 \leftrightarrow P_3.
\end{equation}

\subsubsection{First-layer potential conservation laws}

In the same way than for the PDE, 
there is a one-to-one explicit relationship between multipliers and conservation laws for the first-layer potential system 
\eqref{1stlayer.ux}--\eqref{1stlayer.ut},
\begin{equation}\label{pot1.1stlayer.conslaw}
(f'(p) p_t - \mu p_{xx} - u_x)Q^p+(c^2 p_x + (\b - \mu) p_{tx}  - u_t)Q^u=D_t T+D_x \Phi
\end{equation}
where $Q=(Q^p,Q^u)$.
The determining equations for multipliers 
using the Euler operator are
\begin{subequations}
\begin{align}
E_p((f'(p) p_t - \mu p_{xx} - u_x)Q^p+(c^2 p_x + (\b - \mu) p_{tx}  - u_t)Q^u)&=0,
\\
E_u((f'(p) p_t - \mu p_{xx} - u_x)Q^p+(c^2 p_x + (\b - \mu) p_{tx}  - u_t)Q^u)&=0.
\end{align}
\end{subequations}
These equations yield the following result for all low-order multipliers.

\begin{proposition}
The low-order multipliers for the first-layer potential system 
\eqref{1stlayer.ux}--\eqref{1stlayer.ut}
with $f''(p)\neq0$ and $\b\neq0$ are
\begin{equation}
Q_1 =(1,0),
\quad
Q_2 =(0,1),
\quad
Q_3=(x,-t) .
\end{equation}
\end{proposition}
Then, using the same procedure than for the PDE,
the following conservation laws are obtained
making $T$ as local as possible in terms of $p$.
\begin{theorem}
The low-order conservation laws admitted by the first-layer potential system
\eqref{1stlayer.ux}--\eqref{1stlayer.ut}
with $f''(p)\neq0$ and $\b\neq0$ are
\begin{align}
& 
T_1 = f(p) 
, 
&&
\Phi_1 = - \mu  p_x - u
,
\\
& 
T_2 = - \partial_x^{-1} (f(p)_t)
, 
&&
\Phi_2 =  c^2 p + \b p_t
,
\\
& 
T_3 = x f(p) + t \partial_x^{-1} (f(p)_t)
, 
&&
\Phi_3 = -c^2 t p  -\b  t p_t - x \partial_x^{-1} (f(p)_t)
,
\end{align} 
where $u=\partial_x^{-1} (f(p)_t) - \mu p_{x}$.
\end{theorem}
It can be seen that $T_2$ and $T_3$ correpond to nonlocal conserved densities
and they are not equivalent to any of the ones for equation \eqref{pde}.

\subsubsection{Second-layer potential point symmetries}

Potential symmetries in characteristic form
\begin{equation}\label{pot1.2ndlayer.X}
\hat\X = P^p\partial_p + P^v\partial_v+P^w\partial_w
\end{equation}
come from the equations
\begin{subequations}
\begin{align}
P^p f'(p) -D_x P^v|_\Esp= 0, 
\quad
&
\mu D_x P^p + P^u- D_t P^v|_\Esp = 0 ,
\label{pot1.2ndlayer.det.symm.1}
\\
c^2 P^p + (\b - \mu - \sigma) D_t P^p - D_t P^w|_\Esp = 0 , 
\quad
&
-\sigma D_x P^p + P^u  - D_x P^w|_\Esp=0 
\label{pot1.2ndlayer.det.symm.2}
\end{align}
\end{subequations}
where $\Esp$ denotes the solution space of equations 
\eqref{2ndlayer.vx}--\eqref{2ndlayer.wx}.

The determining equations \eqref{pot1.2ndlayer.det.symm.1}--\eqref{pot1.2ndlayer.det.symm.2}
can be straightforwardly solved giving all point symmetries
of the second-layer potential system 
\eqref{2ndlayer.vx}--\eqref{2ndlayer.wx} with
\begin{equation}
P^p=\eta^p-\tau p_t-\xi p_x,
\quad
P^v=\eta^v-\tau v_t-\xi v_x,
\quad
P^w=\eta^w-\tau w_t-\xi w_x
\end{equation}
being their characteristic form.

\begin{theorem}\label{thm:pot1.2ndlayer.symms}
The point symmetries of the second-layer potential system 
\eqref{2ndlayer.vx}--\eqref{2ndlayer.wx}
with $f''(p)\neq0$ and $\b\neq0$ are comprised by
a time-translation, a space-translation, a $v$-translation, a $w$-translation, and a $u$-translation combined with a time-dependent shift in $v$ and space-dependent shift in $w$.
Their characteristic forms are
\begin{equation}
\begin{aligned}
	&
	P_1^v = -v_t,
	\quad
	P_2^v = -v_x,
	\quad
	P_3^v = 1,
	\quad
	P_4^v = 0,
	\quad
	P_5^v = t,
	\\
	&
	P_1^w = -w_t,
	\quad
	P_2^w = -w_x,
	\quad
	P_3^w = 0,
	\quad
	P_4^w = 1,
	\quad
	P_5^w = x.
\end{aligned}
\end{equation}
For
$f(p)=k(p+p_0)^{1+q}$,
an additional scaling in $x$, $p$, $u$, $v$ and time-dependent shift in $w$
\begin{equation}
\begin{aligned}
	&
	P_6^v = -(q+2) v -q x v_x,
	\\
	&
	P_6^w = -2(c^2 p_0 t +w) -q x w_x
\end{aligned}
\end{equation}
\end{theorem}

Application of \eqref{pot1.2ndlayer.X} to equations \eqref{2ndlayer.vx}--\eqref{2ndlayer.wx} gives
$f(p)=D_x P^v$,
$\mu D_x P^p + P^u= D_t P^v$,
$c^2 P^p+(\b-\mu-\sigma) D_t P^p = D_t P^w$,
and 
$- \sigma D_x P^p + P^u = D_x P^w$.

On the one hand,
solving for $P^w$,
the third equation gives
$P^w = (c^2 D_t^{-1} +\b-\mu-\sigma) P^p$. 
Operating the second equation minus the fourth equation yields
 $(\mu+\sigma)D_x P^p = D_t P^v - D_x P^w$
and substituting in this expression the previous relation for $P^w$ and $P^p$, we have
$(\mu+\sigma)D_x P^p = D_t P^v - (c^2 D_t^{-1}+\b-\mu-\sigma) D_x P^p$.
Then, solving it for $P^p$ leads to
$ ( ( \mu+\sigma ) D_x +(c^2 D_t^{-1} + \b-\mu-\sigma) D_x )P^p = 
D_t P^v$
and inverting the operator,
\begin{equation}\label{2ndlayer.symm.project.v}
P^p = ( c^2 D_t^{-1}+\b D_x  )^{-1} ( D_t P^v ).
\end{equation}

On the other hand,
from the previous relation between $P^w$ and $P^p$ we have
\begin{equation}\label{2ndlayer.symm.project.w}
P^p=(c^2+(\b-\mu-\sigma) D_t)^{-1} (D_t P^w).
\end{equation}
Hence,
the projections from the second-layer layer potential system \eqref{2ndlayer.vx}--\eqref{2ndlayer.wx}
to the PDE \eqref{pde} are \eqref{2ndlayer.symm.project.v}
and \eqref{2ndlayer.symm.project.w}. 


Substitution of the six symmetries from Theorem \ref{thm:pot1.2ndlayer.symms} into projections \eqref{2ndlayer.symm.project.v} and \eqref{2ndlayer.symm.project.w} yields
\begin{equation}
\begin{aligned}
&
P_1^p = -p_t ,
\quad
P_2^p = -p_x,
\quad
P_3^p = 0,
\quad
P_4^p = 0,
\quad
P_5^p = 0,
\quad
P_6^p = -2(p+p_0)- q x p_x.
\end{aligned}
\end{equation}
The first two symmetries and the last symmetry are inherited from equation \eqref{pde} via the projections \eqref{2ndlayer.symm.project.v} and \eqref{2ndlayer.symm.project.w}.
The other symmetries only exist for the second-layer potential system \eqref{2ndlayer.vx}--\eqref{2ndlayer.wx}. 

The correspondence between the inherited point symmetries 
of the potential system \eqref{2ndlayer.vx}--\eqref{2ndlayer.wx} and the point symmetries of equation \eqref{pde}
is
\begin{equation}
(P^v_1,P^w_1) \leftrightarrow P_1;
\quad
(P^v_2,P^w_2) \leftrightarrow P_2;
\quad
(P^v_6,P^w_6) \leftrightarrow P_3.
\end{equation}

\subsubsection{Second-layer potential conservation laws}

The correspondence between multipliers and conservation laws
for the second-layer potential system \eqref{2ndlayer.vx}--\eqref{2ndlayer.wx} is
\begin{equation}\label{pot1.2ndlayer.conslaw}
\begin{aligned}
(f(p)- v_x)Q^p+( \mu p_x  - u - v_t)Q^u+(c^2 p + (\b - \mu - \sigma) p_t - w_t)Q^v+&
\\
(-\sigma p_x + u - w_x)Q^w &=D_t T+D_x \Phi
\end{aligned}
\end{equation}
where $Q=(Q^p,Q^u,Q^v,Q^w)$.
The determining equations for all low-order multipliers are
\begin{subequations}
\begin{align}
E_p((f(p)- v_x)Q^p+( \mu p_x  - u - v_t)Q^u+(c^2 p + (\b - \mu - \sigma) p_t - w_t)Q^v+&
\\
(-\sigma p_x + u - w_x)Q^w)&=0,
\\
E_u((f(p)- v_x)Q^p+( \mu p_x  - u - v_t)Q^u+(c^2 p + (\b - \mu - \sigma) p_t - w_t)Q^v+&
\\
(-\sigma p_x + u - w_x)Q^w)&=0,
\\
E_v((f(p)- v_x)Q^p+( \mu p_x  - u - v_t)Q^u+(c^2 p + (\b - \mu - \sigma) p_t - w_t)Q^v+&
\\
(-\sigma p_x + u - w_x)Q^w)&=0,
\\
E_w((f(p)- v_x)Q^p+( \mu p_x  - u - v_t)Q^u+(c^2 p + (\b - \mu - \sigma) p_t - w_t)Q^v+&
\\
(-\sigma p_x + u - w_x)Q^w)&=0.
\end{align}
\end{subequations}

\begin{proposition}
The low-order multiplier for the second-layer potential system
\eqref{2ndlayer.vx}--\eqref{2ndlayer.wx} with 
$f''(p)\neq0$ and $\b\neq0$ is
\begin{equation}
Q =(0,-1,0,1) .
\end{equation}
\end{proposition}

Thus, we get to the following result for conservation laws.
\begin{theorem}
The low-order conservation law admitted by the second-layer potential system
\eqref{2ndlayer.vx}--\eqref{2ndlayer.wx} 
with $f''(p)\neq0$ and $\b\neq0$ is
\begin{align}
& 
T= v, 
&&
\Phi= - (\mu + \sigma) p - w  .
\end{align} 
\end{theorem}
$T$ corresponds to
a nonlocal conserved density involving
$v=\partial_x^{-1} (f(p))$,
arising from equation \eqref{2ndlayer.vx}.
Therefore,
this conservation law is not locally equivalent to any of the ones for equation \eqref{pde}.

\subsection{Second potential system}

Similarly to the situation for the first potential system,
starting from the generalized Westervelt equation \eqref{pde},
a potential $v(t,x)$ can be introduced yielding 
a first-layer potential system,
\begin{subequations}
\begin{align}
f'(p) p_t - \b p_{xx}   & = v_{xx},
\label{pot2.1stlayer.vxx}
\\
c^2 p  & = v_t .
\label{pot2.1stlayer.vt}
\end{align}
\end{subequations}
Both equations lead to a potential equation,
\begin{equation}
(f(\tfrac{1}{c^2} v_t) - \tfrac{\b}{c^2} v_{xx})_t = v_{xx}.
\end{equation}
From this single potential equation,
a new potential $w(t,x)$ can be introduced 
yielding a second-layer potential system,
\begin{subequations}
\begin{align}
f(\tfrac{1}{c^2} v_t)- \tfrac{\b}{c^2} v_{xx}   & = w_{xx},
\label{pot2.2ndlayer.wxx}
\\
v  & = w_t .
\label{pot2.2ndlayer.wt}
\end{align}
\end{subequations}
Both equations also yield a potential equation,
\begin{equation}
f(\tfrac{1}{c^2} w_{tt}) - \tfrac{\b}{c^2} w_{txx} = w_{xx}.
\end{equation}

\subsubsection{First-layer potential point symmetries}

Potential symmetries in characteristic form
\begin{equation}\label{pot2.1stlayer.X}
\hat\X = P^p\partial_p + P^v\partial_v
\end{equation}
arise from the equations
\begin{subequations}
\begin{align}
D_t (P^p f'(p)) -\b D_x^2 P^p - D_x^2 P^v|_\Esp =0,
\label{pot2.1stlayer.det.symm.1}
\\
c^2 P^p - D_t P^v|_\Esp = 0
\label{pot2.1stlayer.det.symm.2}
\end{align}
\end{subequations}
where $\Esp$ denotes the solution space of equations 
\eqref{pot2.1stlayer.vxx}--\eqref{pot2.1stlayer.vt}.

The determining equations \eqref{pot2.1stlayer.det.symm.1}--\eqref{pot2.1stlayer.det.symm.2}
lead to all point symmetries of the first-layer potential system
\eqref{pot2.1stlayer.vxx}--\eqref{pot2.1stlayer.vt}
with
\begin{subequations}
\begin{align}
P^p&=\eta^p(t,x,p,v)-\tau(t,x,p,v)p_t-\xi(t,x,p,v)p_x,
\\
P^v&=\eta^v(t,x,p,v)-\tau(t,x,p,v)v_t-\xi(t,x,p,v)v_x
\end{align}
\end{subequations}
as their characteristic form.

\begin{theorem}\label{thm:pot2.1stlayer.symms}
The point symmetries of the first-layer potential system \eqref{pot2.1stlayer.vxx}--\eqref{pot2.1stlayer.vt} with $f''(p)\neq0$ and $\b\neq0$ are comprised by a time-translation, a space-translation, a $v$-translation, and a space-dependent shift in $v$.
Their characteristic forms are
\begin{equation}
P_1^v = -v_t ,
\quad
P_2^v = -v_x,
\quad
P_3^v = 1,
\quad
P_4^v = x.
\end{equation}
For
$f(p)=k(p+p_0)^{1+q}$,
an additional scaling in $x$, $p$, $v$ and time-dependent shift in $v$
\begin{equation}
	P_5^v = -2(c^2 p_0 t+v) - q x v_x.
\end{equation}
For
$f(p)=\tfrac{k}{(p+p_0)^3}$,
an additional non-rigid scaling in $x$, $p$, $v$ and time-dependent shift in $v$
\begin{equation}
	P_6^v = x (c^2 p_0 t + v) - x^2 v_x.
\end{equation}
\end{theorem}

Application of \eqref{pot2.1stlayer.X} to equation \eqref{pot2.1stlayer.vt}
directly gives
\begin{equation}\label{pot2.1stlayer.projection}
P^p = D_t P^v / c^2
\end{equation}
arising from $p = v_t / c^2$. 
This expression is the projection formula from
the first-layer potential system \eqref{pot2.1stlayer.vxx}--\eqref{pot2.1stlayer.vt}
to the PDE \eqref{pde}.


Substitution of the six symmetries from Theorem \ref{thm:pot2.1stlayer.symms} into projection \eqref{pot2.1stlayer.projection} yields
\begin{equation}
\begin{aligned}
&
P_1^p = -p_t ,
\quad
P_2^p = -p_x,
\quad
P_3^p = 0,
\quad
P_4^p = 0,
\\
&
P_5^p = -2(p+p_0) - q x p_x, 
\\
&
P_6^p = x(p + p_0) - x^2 p_x.
\end{aligned}
\end{equation}
The first two symmetries and the last two symmetries are inherited from equation \eqref{pde} via the projection \eqref{pot2.1stlayer.projection}.
The other symmetries only exist for the first-layer potential system \eqref{pot2.1stlayer.vxx}--\eqref{pot2.1stlayer.vt}. 

The correspondence between the inherited point symmetries 
of the potential system \eqref{pot2.1stlayer.vxx}--\eqref{pot2.1stlayer.vt} and the point symmetries of equation \eqref{pde}
is 
\begin{equation}
P^v_1 \leftrightarrow P_1;
\quad
P^v_2 \leftrightarrow P_2;
\quad
P^v_5 \leftrightarrow P_3;
\quad
P^v_6 \leftrightarrow P_4.
\end{equation}

\subsubsection{First-layer potential conservation laws}
The relation between multipliers and conservation laws
for the first-layer potential system \eqref{pot2.1stlayer.vxx}--\eqref{pot2.1stlayer.vt} is
\begin{equation}\label{pot2.1stlayer.conlaw}
(f'(p) p_t - \b p_{xx} - v_{xx})Q^p+(c^2 p - v_t)Q^v=D_t T+D_x \Phi
\end{equation}
where $Q=(Q^p,Q^v)$.
The determining equations to obtain all low-order multipliers are
\begin{subequations}
\begin{align}
E_p((f'(p) p_t - \b p_{xx} - v_{xx})Q^p+(c^2 p  - v_t)Q^v)&=0,
\\
E_v((f'(p) p_t - \b p_{xx} - v_{xx})Q^p+(c^2 p  - v_t)Q^v)&=0.
\end{align}
\end{subequations}

\begin{proposition}
The low-order multipliers for the first-layer potential system 
\eqref{pot2.1stlayer.vxx}--\eqref{pot2.1stlayer.vt}
with $f''(p)\neq0$ and $\b\neq0$ are
\begin{equation}
Q_1 =(1,0),
\quad
Q_2 =(x,0) .
\end{equation}
\end{proposition}

Then, the corresponding conservation laws are determined.

\begin{theorem}
The low-order conservation laws admitted by the first-layer potential system \eqref{pot2.1stlayer.vxx}--\eqref{pot2.1stlayer.vt}
with $f''(p)\neq0$ and $\b\neq0$ are
\begin{align}
& 
T_1 = f(p) , 
&&
\Phi_1 = - \b p_x - v_x  ,
\\
& 
T_2 = x f(p) , 
&&
\Phi_2 =  \b (p -x p_x) + v - x v_x  .
\end{align} 
\end{theorem}
It can be seen that the fluxes $\Phi_1$ and $\Phi_2$ are nonlocal,
involving $v=c^2 \partial_t^{-1} p$. 
Hence,
these conservation laws are not locally equivalent to any of the ones for equation \eqref{pde}.

\subsubsection{Second-layer potential point symmetries}

Potential symmetries in characteristic form
\begin{equation}\label{pot2.2ndlayer.X}
\hat\X = P^v\partial_v + P^w\partial_w
\end{equation}
come from the equations
\begin{subequations}
\begin{align}
\tfrac{1}{c^2} f'(\tfrac{1}{c^2} v_t) D_t P^v - \tfrac{\b}{c^2} D_x^2 P^v - D_x^2 P^w|_\Esp   & = 0,
\label{pot2.2ndlayer.det.symm.1}
\\
P^v - D_t P^w |_\Esp & = 0 
\label{pot2.2ndlayer.det.symm.2}
\end{align}
\end{subequations}
where $\Esp$ denotes the solution space of equations
\eqref{pot2.2ndlayer.wxx}--\eqref{pot2.2ndlayer.wt}.

The determinining equations \eqref{pot2.2ndlayer.det.symm.1}--\eqref{pot2.2ndlayer.det.symm.2}
can be solved obtaining all point symmetries of the second-layer potential system \eqref{pot2.2ndlayer.wxx}--\eqref{pot2.2ndlayer.wt} with
\begin{subequations}
\begin{align}
P^v=\eta^v(t,x,v,w)-\tau(t,x,v,w)v_t-\xi(t,x,v,w)v_x,
\\
P^w=\eta^w(t,x,v,w)-\tau(t,x,v,w)w_t-\xi(t,x,v,w)w_x
\end{align}
\end{subequations}
being their characteristic form.

\begin{theorem}\label{thm:pot2.2ndlayer.symms}
The point symmetries of the second-layer potential system \eqref{pot2.2ndlayer.wxx}--\eqref{pot2.2ndlayer.wt} with $f''(\tfrac{1}{c^2}v_t)\neq0$ and $\b\neq0$ are comprised by a time-translation, a space-translation, a $w$-translation, a space-dependent shift in $w$, a $v$-translation combined with a time-dependent shift in $w$, and a space-dependent shift in $v$ combined with a time- and space-dependent shift in $w$.
Their characteristic forms are
\begin{equation}
	P_1^w = -w_t ,
	\quad
	P_2^w = -w_x,
	\quad
	P_3^w = 1,
	\quad
	P_4^w = x,
	\quad
	P_5^w = t,
	\quad
	P_6^w = t x.
\end{equation}
For
$f(\tfrac{1}{c^2}v_t)=k(\tfrac{1}{c^2}v_t+v_0)^{1+q}$,
an additional scaling in $x$, $v$, $w$ and time-dependent shift in $v$ and $w$
\begin{equation}
	P_7^w = - (v_0 t^2 + 2w) - q x w_x .
\end{equation}
For
$f(\tfrac{1}{c^2}v_t)=\tfrac{k}{(\tfrac{1}{c^2}v_t+v_0)^3}$,
an additional non-rigid scaling in $x$, $v$, $w$ and time-dependent shift in $v$ and $w$
\begin{equation}
	P_8^w = x(v_0 t^2 +2w) -2 x^2 w_x .
\end{equation}
\end{theorem}

Application of \eqref{pot2.2ndlayer.X} to \eqref{pot2.2ndlayer.wt}
directly gives
\begin{equation}\label{pot2.2ndlayer.projection}
P^p = D_t^2 P^w / c^2
\end{equation}
arising from $p = v_t / c^2 = w_{tt} / c^2$. 
This expression is the projection formula from
the second-layer potential system \eqref{pot2.2ndlayer.wxx}-\eqref{pot2.2ndlayer.wt}
to the PDE \eqref{pde}.


Substitution of the eight symmetries from Theorem \ref{thm:pot2.2ndlayer.symms} into projection \eqref{pot2.2ndlayer.projection} yields
\begin{equation}
\begin{aligned}
&
P_1^p = -p_t ,
\quad
P_2^p = -p_x,
\quad
P_3^p = 0,
\quad
P_4^p = 0,
\quad
P_5^p = 0,
\quad
P_6^p = 0,
\\
&
P_7^p = -2 (\tfrac{v_0}{c^2} + p ) - q x p_x, 
\\
&
P_8^p = 2x (\tfrac{v_0}{c^2} + p) - 2x^2 p_x.
\end{aligned}
\end{equation}
The first two symmetries and the last two symmetries are inherited from equation \eqref{pde} via the projection \eqref{pot2.2ndlayer.projection}.
The other symmetries only exist for the second-layer potential system \eqref{pot2.2ndlayer.wxx}--\eqref{pot2.2ndlayer.wt}. 

The correspondence between the inherited point symmetries 
of the potential system \eqref{pot2.2ndlayer.wxx}--\eqref{pot2.2ndlayer.wt} and the point symmetries of equation \eqref{pde}
is 
\begin{equation}
P^w_1 \leftrightarrow P_1;
\quad
P^w_2 \leftrightarrow P_2;
\quad
P^w_7 \leftrightarrow P_3;
\quad
P^w_8 \leftrightarrow P_4.
\end{equation}

\subsubsection{Second-layer potential conservation laws}

In the same way, 
the correspondence between multipliers and conservation laws
for the second-layer potential system 
\eqref{pot2.2ndlayer.wxx}--\eqref{pot2.2ndlayer.wt}
is
\begin{equation}\label{pot2.2ndlayer.conslaw}
\begin{aligned}
(f(\tfrac{1}{c^2}v_t)- \tfrac{\b}{c^2}v_{xx}-w_{xx})Q^v+( \mu p_x  - u - v_t)Q^w =D_t T+D_x \Phi
\end{aligned}
\end{equation}
where $Q=(Q^v,Q^w)$.
The determining equations for all low-order multipliers are
\begin{subequations}\label{pot2.2ndlayer.det.conslaw}
\begin{align}
E_v((f(\tfrac{1}{c^2}v_t)- \tfrac{\b}{c^2}v_{xx}-w_{xx})Q^v+( \mu p_x  - u - v_t)Q^w)&=0,
\\
E_w((f(\tfrac{1}{c^2}v_t)- \tfrac{\b}{c^2}v_{xx}-w_{xx})Q^v+( \mu p_x  - u - v_t)Q^w)&=0.
\end{align}
\end{subequations}

\begin{proposition}
No low-order conservation laws are admitted by the second-layer potential system \eqref{pot2.2ndlayer.wxx}--\eqref{pot2.2ndlayer.wt}
with $f''(\tfrac{1}{c^2}v_t)\neq0$ and $\b\neq0$. 
\end{proposition}

\section{Travelling wave solutions}
\label{sec:tws}

The purpose of this section is to apply symmetry reductions
using the point symmetries previously obtained
to transform the original partial differential equation
into ordinary differential equations.
Afterwards,
the reduced equations can be solved leading to
invariant solutions of the governing equation.

Specifically, 
symmetry reductions relevant to the goal of this work with physical interest
are only translations, which yield travelling wave solutions.
The other symmetries correspond to cases where $f(p)$ does not follow the Westervelt's equation structure.

Consider
\begin{equation}\label{tw}
p = U(\xi),
\quad
\xi = x- \nu t.
\end{equation}
Equation (\ref{pde}) with $f(p) =p +h(p)$ being $h(p)$ a nonlinear function becomes
\begin{equation}\label{pde2}
(p+	h(p))_{tt} - \b p_{xxt} = c^2 p_{xx}.
\end{equation}
Substitution of (\ref{tw}) into equation (\ref{pde2}) yields the third-order ODE
\begin{equation}\label{ode}
\b \nu U'''+(\nu^2 -c^2)	U''+\nu^2 ( h'(U) U''+h''(U) U'^2)=0.
\end{equation}
Integrating twice with respect to the independent variable $\xi$ we have that ODE (\ref{ode})
reduces to a first-order ODE,
\begin{equation}\label{ode1}
\b \nu U'+(\nu^2 -c^2)	U+\nu^2 h(U)+C_1 \xi+C_0=0,
\end{equation}
which can be expressed as a quadrature by setting $C_1=C_0=0$,
\begin{equation}
\int_{U_0}^U \frac{\b\nu}{ (\nu^2-c^2)U+\nu^2 h( U )}dU
=-(\xi-\xi_0).
\end{equation}

We now will pursue a physically interesting solution that describes a shock wave.

\subsection{Shock wave solutions}
In general, 
nonlinear PDEs studied in the context of wave phenomena
lead to various forms of wave solutions
such as solitary waves, shock waves, cuspons, peakons
and other types.
In this context shock wave appear. 
They represent a sharp discontinuity of the parameters
that describe the media
and they are characterized by a wave amplitude that shows an exponential change between two asymptotically constant values. 
Unlike solitons where the energy is a conserved quantity,
shock waves dissipate energy with distance.
It is important to point out that shock wave solutions
correspond to discontinous weak solutions.


From the first-order ODE \eqref{ode1}, 
if we assume $h(U)=\kappa U^n$, with $\kappa>0$, and $C_0 = C_1 = 0$,
then the first-order ODE becomes
\begin{equation}\label{ode2}
\b \nu U'+\nu^2 \kappa U^n+ (\nu^2 -c^2)	U=0.
\end{equation} 
This equation is a Bernoulli type equation and 
can be straightforwardly solved,
explicitly obtaining the following solution,
\begin{equation}\label{sol0}
U(\xi)=\left(\frac{c^2-\nu^2}{U_0(c^2-\nu^2) e^{-\tfrac{(c^2-\nu^2)(n-1)\xi}{\beta \nu}}+\kappa \nu^2}\right)^{\tfrac{1}{n-1}}
\end{equation}
where $n>1$ and $U_0$ is an arbitrary constant.

\begin{figure}[h]
\centering
\includegraphics[width=0.48\textwidth]{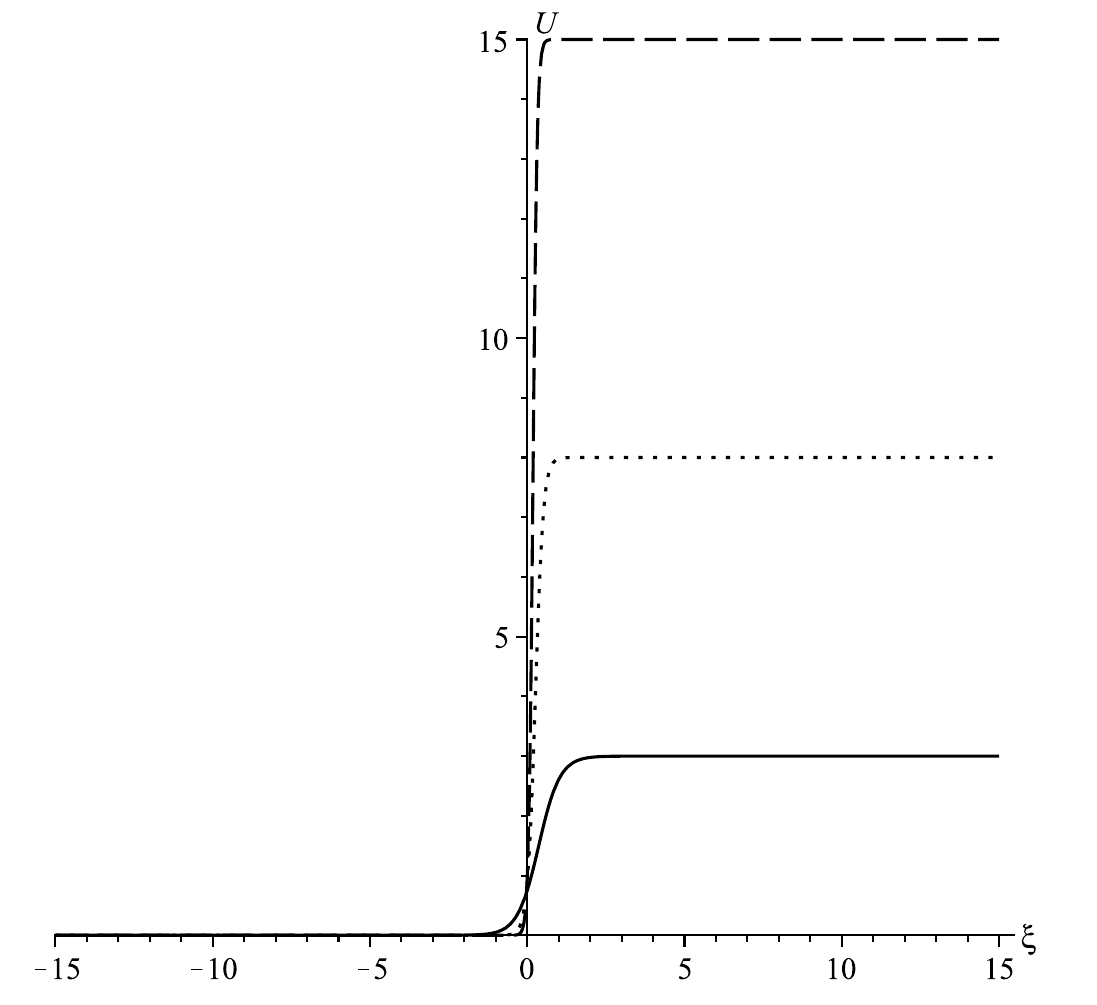}
\quad
\includegraphics[width=0.48\textwidth]{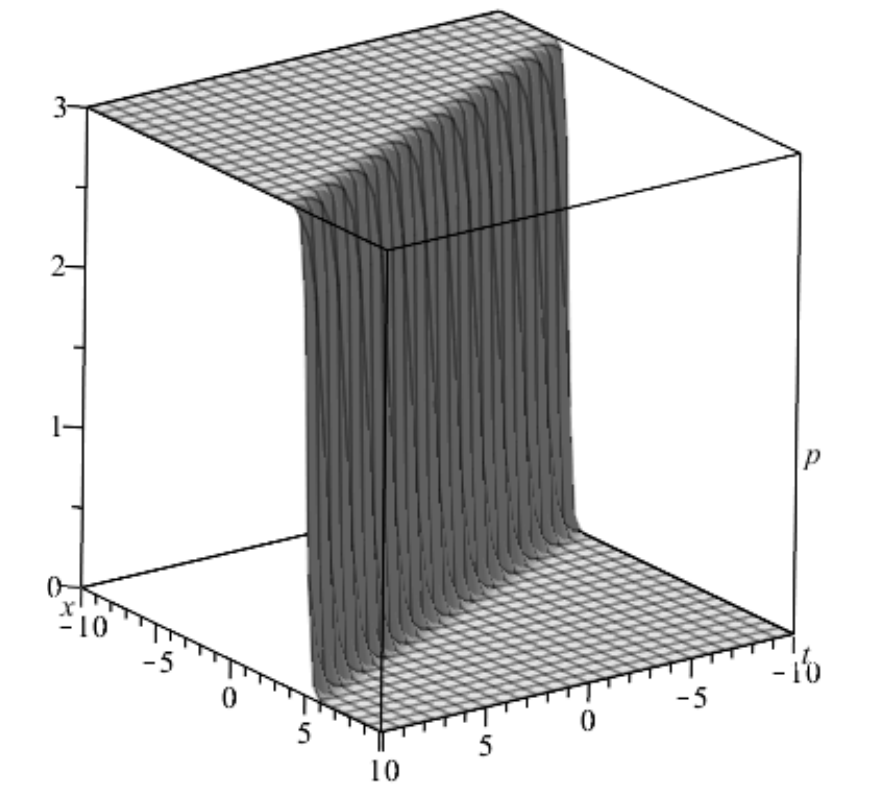}
\caption{Shock wave profile \eqref{sol1} (left) for $c=2$ (solid), $c=3$ (dot), $c=4$ (dash). 3D shock wave space-time \eqref{sol1} (right) for $c=2$ and $\nu=2$. The parameters used are $U_0=\nu=\b=\kappa=1$.}
\label{fig:shock}
\end{figure}

\begin{figure}[h]
	\centering
	\includegraphics[width=0.48\textwidth]{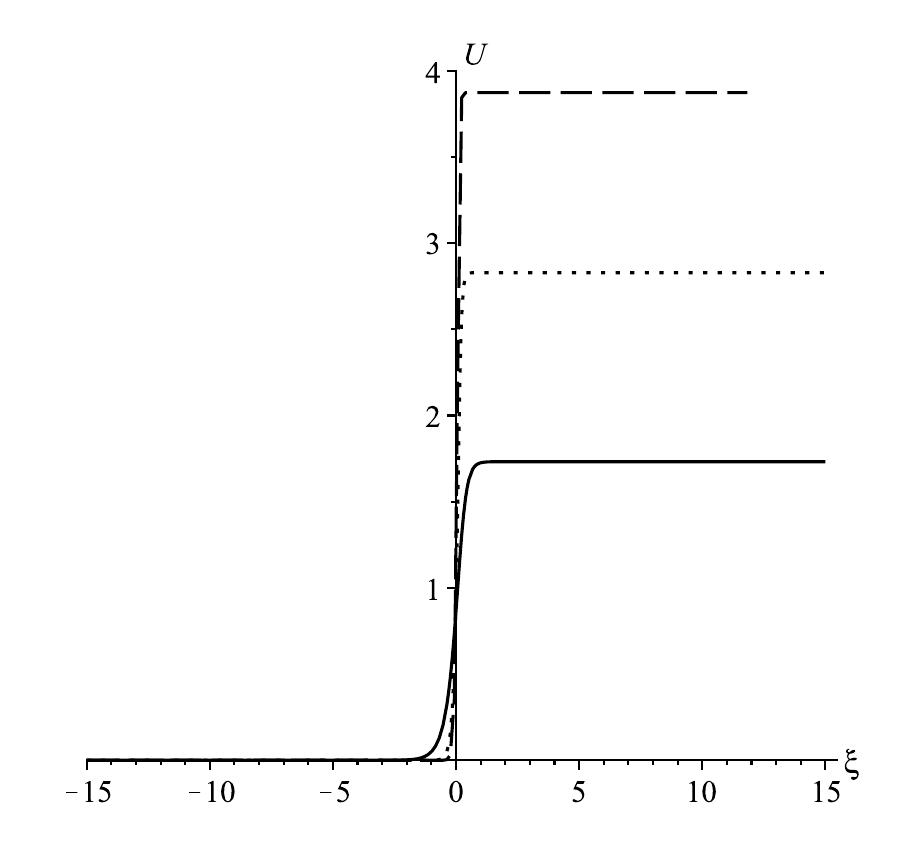}
	\quad
	\includegraphics[width=0.48\textwidth]{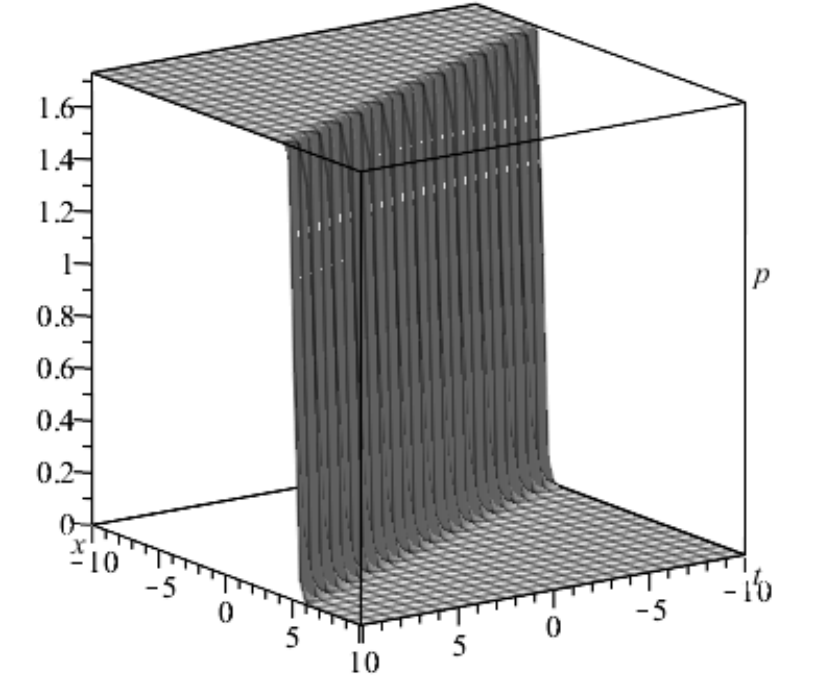}
	\caption{Shock wave profile \eqref{sol0} (left) for $c=2$ (solid), $c=3$ (dot), $c=4$ (dash). 3D shock wave space-time \eqref{sol0} (right) for $c=2$ and $\nu=2$. The parameters used are $n=3$ and $U_0=\nu=\b=\kappa=1$.}
	\label{fig:shock2}
\end{figure}

Consider the physical situation with $n=2$ corresponding to the original Westervelt's equation,
\begin{equation}\label{sol1}
U(\xi)=\frac{c^2-\nu^2}{U_0(c^2-\nu^2) e^{-\frac{(c^2-\nu^2)\xi}{\beta \nu}}+\kappa \nu^2}.
\end{equation}
If we assume $U_0>0$ and $c^2-\nu^2>0$,
then this solution is non-singular. 
As $\xi \rightarrow -\infty$, we then see that $U' \rightarrow 0$ and $U  \rightarrow 0$, while as $\xi  \rightarrow +\infty$, we see that $U' \rightarrow 0$ and $U \rightarrow \frac{c^2-\nu^2}{\kappa \nu^2}$.
Thus, this solution is a shock wave.
See Figure \ref{fig:shock} for a plot of the shock profile
and of the 3D shock wave.
Nevertheless,
the same qualitative feature is expected to retain for $n=3$ (see Figure \ref{fig:shock2}).

\section{Conclusions}
\label{sec:conclusions}

The present work provides symmetries and conservation laws
for a generalized Westervelt equation \eqref{pde} in one dimension.
A classification of the point symmetries admitted by this equation is presented
with their physical meaning and their commutator structure.

The generalized Westervelt equation \eqref{pde} is found to possess
local low-order conservation laws, 
which describe conserved quantitites related to net mass for sound waves. 

From the main PDE \eqref{pde},
two potential systems are derived by the introduction of first- and second-layer potentials. 
The point symmetries of these systems are studied
yielding symmetries inherited from equation \eqref{pde} via projections 
and also symmetries that only exist for the potential systems.
In addition,
the multiplier method is also applied to the potential systems 
leading to nonlocal conservation laws.
 
For equation \eqref{pde}, 
the physically meaningful symmetry reductions arise from translations
and yield travelling wave solutions.
The general solution for $f(p)=p+h(p)$
being $h(p)$ a nonlinear function
is presented as a quadrature. 
Then, 
we consider the physically interesting case 
with $f(p)=p+\kappa p^n$, $\kappa>0$, $n>1$,
leading to shock wave solutions,
whose behaviour is analyzed.

\section*{Acknowledgements}
The authors are supported by the \textit{Junta de Andaluc\'ia} research group FQM-201. 
We gratefully acknowledge Dr. Stephen Anco from Brock University
for his expert guidance and help during his visit to the University of Cadiz.


\begin{thebibliography}{99}


\bibitem{HamBla-book}
M. Hamilton and D. Blackstock,
{\em Nonlinear Acoustics},
Graduate Studies in Mathematics, Academic Press, 1998.

\bibitem{MenCaiLiZHoNiuZhe}
L. Meng, F. Cai, F. Li, W. Zhou, L. Niu, H. Zheng, 
Acoustic tweezers, 
J. Phys. D: Appl. Phys. 52 (2019), 273001.




\bibitem{Sza}
T.L. Szabo,
{\em Diagnostic ultrasound imaging: inside out},
Academic Press, 2004.

\bibitem{GueMarEglAliSer}
A.G. Guex, N. Di Marzio, D. Eglin, M. Alini, T. Serra, 
The waves that make the pattern: a review on acoustic manipulation in
biomedical research, 
Mater. Today Bio 10 (2021), 100110.


\bibitem{Lur}
X. Lurton,
{\em An Introduction to Underwater Acoustics: Principles and Applications}, Lurton (2nd ed.)
Springer Praxis: London, 2010. 

\bibitem{GanYanKam}
W.S. Gan, J. Yang, T. Kamakura,
A review of parametric acoustic array in air, 
Appl. Acoustics 73(12) (2012), 1211--1219,

\bibitem{Cam}
D.M. Campbell,
Nonlinear dynamics of musical reed and brass wind instruments,
Contemp. Phys. 40(6) (1999), 415--431.

\bibitem{MyePylGilCamChiLog}
A. Myers, R.W. Pyle Jr, J. Gilbert, D.M. Campbell, J.P. Chick, S. Logie,
Effects of nonlinear sound propagation on the characteristic timbres of brass instruments,
J. Acoust. Soc. Am. 131(1) (2012), 678--688.

\bibitem{MasLor}
T.J. Mason, J.P. Lorimer,
{\em Applied sonochemistry: Uses of power ultra-sound in chemistry and processing},
Wiley-VCH: Weinheim (2002).

\bibitem{MaeSev}
R. Gr. Maev, F. Seviaryn, 
Applications of non-linear acoustics for quality control and material characterization, 
J. Appl. Phys. 132 (2022), 161101.








\bibitem{Wes}
P.J. Westervelt,
Parametric acoustic array,
J. Acoustic Soc. Amer. 35 (1963), 535--537.

\bibitem{Tar}
G. Taraldsen,
Generalized Westervelt equation, 
J. Acoust. Soc. Am. 109(4) (2001), 1329-1333.

\bibitem{Jor}
P.M. Jordan,
A survey of weakly-nonlinear acoustic models: 1910-2009,
Mechanics Res. Commun. 73 (2016) 127--139.

\bibitem{EoS}
T. Feroze, A.A. Siddiqui,
Charged anisotropic matter with quadratic equation of state,
Gen. Relativ. Gravit. 43 (2011) 1025--1035.

\bibitem{AncWes}
S.C. Anco, A.P. Márquez, T.M. Garrido, M.L. Gandarias, 
Symmetry analysis and hidden variational structure of Westervelt's equation in nonlinear acoustics, 
Commun. Nonlinear Sci.
124 (2023), 107315.

\bibitem{Anc2022a}
S.C. Anco, 
Symmetry actions and brackets for adjoint-symmetries. II: Physical examples, 
European Journal of Applied Mathematics 34(5) (2023) 974--997.



\bibitem{Artur}
A. Sergyeyev, 
Complete description of local conservation laws for generalized dissipative Westervelt equation, 
Qual. Theory Dyn. Syst. 23 (2024),
209. 


\bibitem{SolSheThi}
M. Solovchuk, T.W. Sheu, M. Thiriet, 
Simulation of nonlinear Westervelt equation for the investigation of acoustic streaming and nonlinear propagation effects, 
J. Acoust. Soc. Am. 134(5) (2013), 3931--3942.

\bibitem{ManSolShe}
M.A. Diaz, M.A. Solovchuk, T.W.H. Sheu, 
A conservative numerical scheme for modeling nonlinear acoustic propagation in thermoviscous homogeneous media, 
J. Computational Phys. 363 (2018), 200--230.




\bibitem{Kal}
B. Kaltenbacher. Mathematics of nonlinear acoustics,
Evolution Equations and Control Theory (EECT), 4 (2015), 447--491. 



\bibitem{Chi}
Y.A. Chirkunov, 
Invariant submodels of the Westervelt model with dissipation, 
Int. J. Non-Linear Mech. 84 (2016), 139--144. 

\bibitem{Ovs-book}
L.V. Ovsiannikov,
{\em Group Analysis of Differential Equations},
Academic Press: New York, 1982.

\bibitem{Olv-book}
P.J. Olver,
{\em Applications of Lie Groups to Differential Equations},
Springer-Verlag, New York, 1993.

\bibitem{BCA-book}
G.W. Bluman, A. Cheviakov, S.C. Anco,
{\em Applications of Symmetry Methods to Partial Differential Equations},
Springer, New York, 2009.

\bibitem{Anc-review}
S.C. Anco,
Generalization of Noether's theorem in modern form to non-variational partial differential equations.
In: Recent progress and Modern Challenges in Applied Mathematics, Modeling and Computational Science
(eds. R. Melnik et al).
Fields Institute Communications, Volume 79, 2017.

\bibitem{AncBlu2002b}
S.C. Anco, G.W. Bluman,
Direct construction method for conservation laws of partial differential equations Part II: General treatment,
Euro. J. Appl. Math. 41 (2002), 567--585.

%
%
%
%
%


\end{thebibliography}
\end{document}